\documentclass[aps,prd,twocolumn,groupedaddress,showpacs,showkeys]{revtex4}

\usepackage{graphicx}
\usepackage{graphicx}
\usepackage{amssymb}
\usepackage{amsmath}
\usepackage{color}
\newcommand{\be}{\begin{equation}}
\newcommand{\ee}{\end{equation}}
 
 \definecolor{BrickRed}{cmyk}{0,0.89,0.94,0.28}
\definecolor{MidnightBlue}{cmyk}{0.98,0.13,0,0.43}
\definecolor{DarkGreen}{rgb}{0,0.7,0.1}

\begin{document}

\title{Casimir-Polder interaction for gently curved surfaces}

\date{\today}

\author{Giuseppe Bimonte$^{1,2}$,
Thorsten Emig$^{3}$, and  Mehran Kardar$^{4}$}  

\affiliation{${ }^{1}$Dipartimento di Scienze Fisiche, Universit\`{a} di
Napoli Federico II, Complesso Universitario
di Monte S. Angelo,  Via Cintia, I-80126 Napoli, Italy\\
${ }^{2}$ INFN Sezione di
Napoli, I-80126 Napoli, Italy\\
${ }^{3}$ Laboratoire de Physique
Th\'eorique et Mod\`eles Statistiques, CNRS UMR 8626, B\^at.~100,
Universit\'e Paris-Sud, 91405 Orsay cedex, France\\
${ }^{4}$ Massachusetts Institute of
Technology, Department of Physics, Cambridge, Massachusetts 02139, USA}

\begin{abstract}
We use a derivative expansion for gently curved surfaces to compute the leading and the next-to-leading curvature corrections to the Casimir-Polder interaction between a polarizable small particle and a non-planar  surface. While our methods apply to any homogeneous and isotropic surface, explicit results are presented here for perfect conductors. We show that the derivative expansion of the Casimir-Polder potential follows from a resummation of its perturbative series, for small in-plane momenta.  We consider the retarded, non-retarded and classical high temperature limits.
\end{abstract}


\maketitle

\section{Introduction}

Quantum and thermal vacuum fluctuations of the electromagnetic field are at the cause of so called dispersion forces between two polarizable  bodies. A particular instance of dispersion interaction is the  Casimir-Polder force~\cite{polder} between a small polarizable particle (like an atom or a molecule) and a nearby material surface. Recent advances in nanotechnology and in the field of ultracold atoms have made possible quite precise measurements
of the Casimir-Polder interaction. (For recent reviews see~\cite{rev1,rev2}.)

There is presently considerable interest in investigating how the Casimir-Polder interaction is affected by the geometrical shape of the surface and several experiments have been recently carried out~\cite{exp1,exp2,exp3,exp4} to probe  dispersion forces between atoms and  microstructured surfaces.    The characteristic non-additivity of dispersion forces make it very difficult to compute the Casimir-Polder interaction for non-planar surfaces in general. Detailed results have been worked out only for a few specific geometries.
The example of a uniaxially  corrugated surface was studied numerically in Ref.~\cite{babette} within a toy scalar field theory, while  rectangular dielectric gratings were considered in Ref.~\cite{dalvit}. In Ref.~\cite{galina} analytical results were obtained for the case of a perfectly conducting cylinder. A perturbative approach is presented in~\cite{messina},  where surfaces with smooth corrugations of any shape, but with small amplitude, were studied. The validity of the latter is  restricted to particle-surface separations that are much larger than the corrugation amplitude. In this paper we present an alternative approach that  becomes exact in the opposite limit of small particle-surface distances. In this limit, the Proximity Force Approximation (PFA)~\cite{deri} can be used to obtain the leading contribution to the Casimir-Polder potential.  Our approach is based on a systematic {\it derivative expansion} of the potential, extending to the Casimir-Polder interaction an analogous expansion successfully used recently~\cite{fosco2,bimonte3,bimonte4} to study the Casimir interaction between two non-planar  surfaces. It has also been applied to other problems involving short range interactions between surfaces, like radiative heat transfer~\cite{golyk} and stray electrostatic forces between conductors~\cite{fosco3}.  From this expansion we could obtain the leading and the next-to-leading curvature corrections to the PFA for the Casimir-Polder interaction.

The paper is organized as follows. In Sec. II we present the derivative expansion for the general case of a dielectric surface. Explicit results are presented for the special case of a perfectly conducting surface. In Sec. III the example of a two-state atom is considered, and we present the potential in two limits: the retarded Casimir-Polder limit and the  non--retarded London limit. In Section IV we  conclude, pointing out some
avenues for further exploration. Finally, in the Appendix we show how the derivative expansion of the potential can be obtained by a  re--summation of the perturbative series to all orders.

\section{Derivative expansion of the Casimir-Polder potential}

\begin{figure} 
\includegraphics [width=.9\columnwidth]{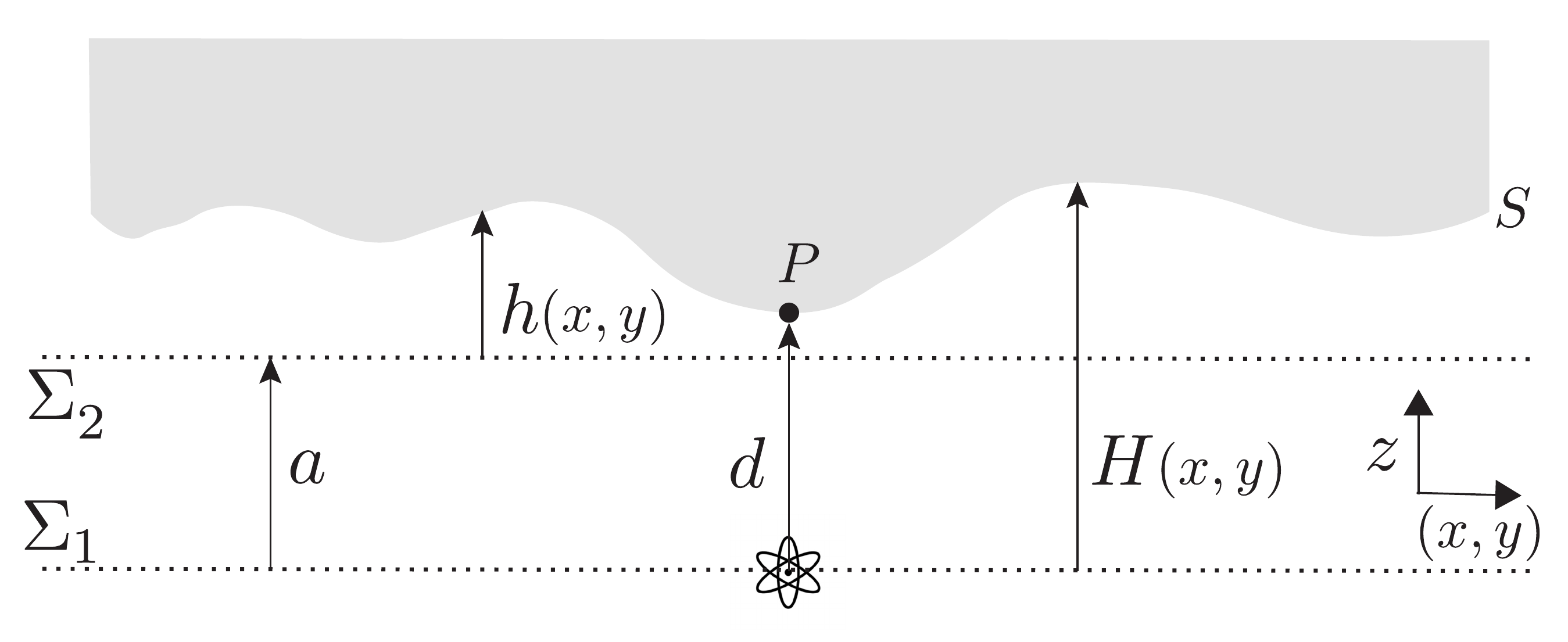} 
\caption{\label{fig1} Coordinates parametrizing a configuration  consisting of  an atom or   nano-particle near a gently curved   surface.}  \end{figure}

Consider a particle (an `atom,'  a molecule, or any polarizable micro-particle) near a dielectric surface $S$. We assume that the particle is small enough (compared to the scale of its
separation $d$ to the surface), such that it can be considered as point-like, with its response to the electromagnetic (em) fields fully described by a dynamic electric dipolar polarizability tensor $\alpha_{\mu \nu}(\omega)$. (We assume for simplicity that the particle has a negligible magnetic polarizability, as is usually the case). Let us denote by $\Sigma_1$ the plane through the atom which is orthogonal to the distance vector (which we take to be the ${\hat {\bf z}}$ axis) connecting the atom to the point $P$ of $S$ closest to the atom.  We assume that the surface $S$ is characterized by a {\it smooth} profile $z=H({\bf x})$, where ${\bf x} =(x,y)$ is the vector spanning $\Sigma_1$, with origin at the atom's position (see Fig.~\ref{fig1}). In what follows greek indices $\mu, \nu, \dots$ label all coordinates $(x,y,z)$, while latin indices $i,j,k \dots$ refer to $(x,y)$ coordinates in the plane $\Sigma_1$. Throughout we adopt the convention that repeated indices are summed over.  

The exact Casimir-Polder potential at finite temperature $T$  is given by the scattering  formula~\cite{sca1,sca2}
\begin{equation}
\label{eq.4}
U= - k_B T \sideset{}{'}\sum_{n=0}^\infty \text{Tr} \, [ {\mathbb T}^{(S)} {\mathbb U} {\mathbb T}^{(A)}  {\mathbb U}  ](\kappa_n)\;.
\end{equation} 
Here ${\mathbb T}^{(S)}$ and ${\mathbb T}^{(A)}$  denote, respectively, the T-operators of the plate $S$ and the atom, evaluated  at the Matsubara wave numbers $\kappa_n=2\pi n k_BT/(\hbar c)$, and the primed sum indicates that the $n=0$ term carries  weight $1/2$.   In a plane-wave basis $|{\bf k},Q \rangle$~\cite{fn3} where ${\bf k}$ is the in-plane wave-vector, and $Q=E,M$ denotes respectively electric (transverse magnetic) and magnetic (transverse electric) modes,
the translation operator $\mathbb U$ in Eq.~(\ref{eq.4}) is diagonal with matrix elements
$e^{-d q}$ where
$q=\sqrt{{ k}^2+ \kappa_n^2}\equiv q(k)$, $k=|{\bf k}|$. The matrix elements of the atom  T-operator in the dipole approximation are:
\be
 { {\cal T}}_{QQ'}^{(A)}({\bf k}, {\bf k}' ) =-\frac{2 \pi \kappa_n^2}{ \sqrt{q q'}}   e^{(+)}_{Q \mu}({\bf k}) \alpha_{\mu \nu}({\rm i} c\kappa_n) e^{(-)}_{Q' \nu}({\bf k}')\;,
\ee
where $q'=q(k')$, ${\bf e}^{(\pm)}_{M}({\bf k})={\hat {\bf z}} \times {\hat {\bf k} }$ and ${\bf e}^{(\pm)}_{E}({\bf k})=-1/\kappa_n ({\rm i} k {\hat {\bf z}} \pm q  {\hat {\bf k} })$, with ${\hat {\bf k} }={\bf k}/k$.
There are no analytical formulae for the elements of the T-operator of a curved plate ${\mathbb T}^{(S)} $, and its computation is in general quite challenging, even numerically. 
We shall demonstrate, however, that for any smooth surface it is possible to compute  the leading curvature corrections to the potential in the experimentally relevant limit of small separations. The key idea is that the Casimir-Polder interaction  falls off rapidly with separation, and it is thus reasonable to expect that the potential $U$ is
mainly determined  by the geometry of the surface  $S$ in a small neighborhood of the point $P$ of $S$ which is {\it closest} to the atom. This physically plausible idea suggests that for small separations $d$ the potential $U$ can be expanded as a series expansion in an increasing number of derivatives of the height profile $H$, evaluated at the atom's position. Up to fourth order, and assuming that the surface is homogeneous and isotropic, the most general expression which is invariant under rotations of the $(x,y)$ coordinates, and that involves at most four derivatives of $H$ (but no first derivatives since $\nabla H ({\bf 0})=0$) can be expressed (at zero temperature, and up to ${\cal O}(d^{-1}$)) in the form 
\begin{widetext}
\begin{align}
\label{derexpa}
U & = -\frac{\hbar c}{d^4} \int_0^{\infty} \frac{d {\xi}}{2 \pi}\left\{\beta^{(0)}_1 \alpha_\perp + \beta^{(0)}_2 \alpha_{zz} + d\times\left[ (\beta^{(2)}_1 \alpha_\perp + \beta^{(2)}_2 \alpha_{zz}) \nabla^2 H+
 \beta^{(2)}_3  \left(\partial_i \partial_j H - \frac{1}{2} \nabla^2 H \delta_{ij}\right) \alpha_{ij}\right]+d^2\times \right. \\ \nonumber
& 
\left[\beta^{(3)}  \alpha_{zi} \partial_i \nabla^2 H \!
 \left.+(\nabla^2 H)^2 (\beta^{(4)}_1 \alpha_\perp\! + \beta^{(4)}_2 \alpha_{zz} )+  (\partial_i \partial_j H)^2 (\beta^{(4)}_3 \alpha_\perp\! + \beta^{(4)}_4 \alpha_{zz} )+ \beta^{(4)}_5
\nabla^2 H \! \left(\partial_i \partial_j H - \frac{1}{2} \nabla^2 H \delta_{ij}\right) \alpha_{ij} \right]\right\}\,,
\end{align}
\end{widetext}
where the Matsubara sum has been replaced by an integral over $\xi=\kappa d$, $\alpha_\perp=\alpha_{xx}+\alpha_{yy}$, and it is understood that all derivatives of $H({\bf x})$ are evaluated at the atom's position i.e. for ${\bf x}={\bf 0}$. The coefficients
$\beta^{(p)}_p$ are dimensionless functions of $\xi$, and of any other dimensionless ratio of frequencies characterizing the material of the surface. The derivative expansion  in Eq.~(\ref{derexpa}) can be formally obtained by a re-summation of the perturbative series for the potential for small in-plane momenta ${\bf k}$ (see Appendix). We note that there are additional terms involving four derivatives of $H$ which, however, yield contributions $\sim 1/d$ (as do terms involving five derivatives of $H$) and are hence  neglected.

As demonstrate	d in the Appendix [see Eqs.~(\ref{Bc}), (\ref{Cc})], the coefficients $\beta^{(p)}_q$ in Eq.~(\ref{derexpa}) can be extracted from  
the perturbative series of the potential $U$, carried to second order in the deformation $h({\bf x})$, which  in turn involves an expansion of the T-operator of the surface $S$ to the same order.
The latter expansion was worked out in Ref.~\cite{voron} for a dielectric surface characterized by a frequency dependent permittvity $\epsilon(\omega)$. It reads
\begin{align}
& { {\cal T}}_{QQ'}^{(S)}({\bf k}, {\bf k}' )=(2 \pi)^2 \delta^{(2)}({\bf k}-{\bf k'})\,\delta_{QQ'}\, r^{(S)}_{Q} (i c\kappa_n,{\bf k})
\nonumber\\
&+ \sqrt{q\,q'}\,\left[-2 \,B_{QQ'}({\bf k}, {\bf k}')\,\tilde{h}({{\bf k}- {\bf k}'})\right. \\
 & \left. + \!\!\int \!\!\frac{d^2 {\bf k}''}{(2 \pi)^2} (B_2)_{QQ'}({\bf k}, {\bf k}';{\bf k}'') \tilde{h}({{\bf k}\!- \!{\bf k}''}) \tilde{h}({{\bf k}''\!-\! {\bf k}'})+\dots \right]\;, \nonumber
\end{align}
where $r^{(S)}_{Q} (i c\kappa_n,{\bf k})$ denotes the familiar Fresnel reflection coeffcient of a flat surface.  Explicit expressions for $B_{Q Q'}({\bf k}, {\bf k}')$ and $(B_2)_{Q Q'}({\bf k}', {\bf k}';{\bf k}'')$ are given in Ref.~\cite{voron}. The computation of the  coefficients $\beta^{(p)}_q$ involves an integral over ${\bf k}$ and ${\bf k'}$ (as it is apparent from Eq.~(\ref{eq.4})) that cannot be performed analytically for a dielectric plate, and has to be  estimated  numerically. In this paper, we shall content ourselves to considering the case of a perfect conductor, in which case the integrals can be performed analytically.  For a perfect conductor, the matrix $B_{QQ'}({\bf k}, {\bf k}')$ takes the simple form
\be
B({\bf k}, {\bf k}')=\left(\begin{array}{cc} \frac{{\hat{\bf k}}\cdot{\hat {\bf k}}'  \kappa_n^2+k k'}{ q q'}& \frac{\kappa_n}{ q} {\hat {\bf z}}\cdot({\hat{\bf k}}\times {\hat {\bf k}}') \\ \frac{\kappa_n}{ q'} {\hat {\bf z}}\cdot({\hat{\bf k}}\times {\hat {\bf k}}') & -{\hat{\bf k}}\cdot{\hat {\bf k}}' \\ \end{array} \right)\;,
\ee 
where the matrix indices $1,2$ correspond to $E,M$ respectively. 
For perfect conductors, the matrix $(B_2)_{QQ'}({\bf k}, {\bf k}';{\bf k}'') $ is simply related to $B$ by
\be
(B_2)({\bf k}, {\bf k}';{\bf k}'') =2 q'' B({\bf k}, {\bf k}'') \sigma_3 B({\bf k}'', {\bf k}')\;,
\ee
where $\sigma_3={\rm diag}(1,-1)$. For perfect conductors the coefficients $\beta^{(p)}_q$ are functions of $\xi$ only, and we list them in  Table~\ref{tab:betas}.

\begin{widetext}
\begin{table*}
\begin{tabular}{|c|c|l|l|}
\hline
p & q & $\times e^{-2\xi}$  & $\times \text{Ei}(2\xi)$ \\
\hline
0 & 1 &  $\frac{1}{8}(1+2 { \xi}+4 { \xi}^2)$ &  $0$\\
 & 2 & $\frac{1}{4}(1+2 { \xi} )$ & $0$\\
2 & 1 & $-\frac{1}{32}(3+6 { \xi} +6 { \xi}^2+4 { \xi}^3 )$ & $-\frac{ { \xi}^4}{4} $\\
 & 2 & $-\frac{1}{16}(1+2 { \xi} -2 { \xi}^2+4 { \xi}^3 )$ & ${ \xi}^2 \left(1-\frac{{ \xi}^2}{2} \right)$\\
 & 3 & $-\frac{1}{32}(3+6 { \xi} +2{ \xi}^2-4 { \xi}^3 )$  & $\frac{ { \xi}^4}{4}$\\
3 & & $\frac{1}{32}(1+2 { \xi} -2{ \xi}^2+4 { \xi}^3 )$ & $-\frac{ { \xi}^2}{4} (2-{ \xi}^2)$\\
4 & 1 & $\frac{1}{384}(3+6 { \xi} +15{ \xi}^2+22 { \xi}^3+2 { \xi}^4-4{ \xi}^5 )$ & $\frac{ { \xi}^4}{48} (6-{ \xi}^2)$\\
 & 2 &  $-\frac{1}{960}(15+542 { \xi} +259{ \xi}^2-546 { \xi}^3-14 { \xi}^4+28{ \xi}^5 )$ & $\frac{ {120 \xi}^4}{7} (20-{ \xi}^2)$\\
 & 3 &  $\frac{1}{192}(15+30 { \xi} -9{ \xi}^2+70 { \xi}^3+2 { \xi}^4-4{ \xi}^5 )$ & $\frac{ { \xi}^4}{24} (18-{ \xi}^2) $\\
 & 4 &  $\frac{1}{480}(45+218 { \xi} -59{ \xi}^2+146 { \xi}^3+14 { \xi}^4-28{ \xi}^5 )$ & $\frac{ { \xi}^4}{60} (40-7{ \xi}^2)$\\
 & 5 &  $\frac{1}{96}(9+18 { \xi} -27{ \xi}^2+50 { \xi}^3-2 { \xi}^4+4{ \xi}^5 )$ & $ { \xi}^4\left(1+\frac{{ \xi}^2}{12}\right)$\\
\hline
\end{tabular}
\caption{\label{tab:betas} The coefficients $\beta^{(p)}_q$ are obtained by multiplying the third column by $e^{-2\xi}$, and adding the fourth column times $\text{Ei}(2\xi)=-\int_{2\xi}^\infty dt \exp(-t)/t$.}
\end{table*}

The geometric significance of the expansion in Eq.~(\ref{derexpa}) becomes more transparent when the $x$ and $y$ axis are chosen to be coincident with the principal directions of $S$ at $P$, in which case the local expansion of $H$ takes the simple form $H=d+x^2/(2 R_1)+ y^2/(2 R_2)+\cdots$ where $R_1$ and $R_2$ are the radii of curvature at $P$.  In this coordinate system, the derivative expansion of $U$ reads
\begin{align}
U & =-\frac{\hbar c}{d^4} \int_0^{\infty} \frac{d \xi}{2 \pi}\left\{\beta^{(0)}_1 \alpha_\perp + \beta^{(0)}_2 \alpha_{zz} + \left(\frac{d}{R_1}+\frac{d}{R_2} \right) (\beta^{(2)}_1 \alpha_\perp + \beta^{(2)}_2 \alpha_{zz})+
 \frac{\beta^{(2)}_3}{2}  \left(\frac{d}{R_1}-\frac{d}{R_2}\right) (\alpha_{xx}-\alpha_{yy}) \right.\nonumber \\
&+ d^2 \beta^{(3)}  \alpha_{zi} \partial_i \left(\frac{1}{R_1}
+\frac{1}{R_2} \right) 
+ \left(\frac{d}{R_1}+\frac{d}{R_2} \right)^2 (\beta^{(4)}_1 \alpha_\perp + \beta^{(4)}_2 \alpha_{zz} )\frac{}{}
\nonumber \\
& \left.\left.
+ \left[\left(\frac{d}{R_1}\right)^2+\left(\frac{d}{R_2} \right)^2 \right]   (\beta^{(4)}_3 \alpha_\perp + \beta^{(4)}_4 \alpha_{zz} )+ \frac{\beta^{(4)}_5}{2} \left[\left(\frac{d}{R_1}\right)^2-\left(\frac{d}{R_2} \right)^2  \right] (\alpha_{xx}-\alpha_{yy}) 
  \right\}\right.\;.\label{derexpa2}
\end{align}
\end{widetext}

\section{Two-state ``atom''}

The $\beta$ coefficients in Eq.~(\ref{derexpa}) are significantly different from zero only
for rescaled  frequencies $\xi \lesssim 1$.  Therefore, for separations small compared to the radii of surface curvature  but $d \gg c/\omega_r$, where $\omega_r$ is the
typical atomic resonance frequency, we can replace $\alpha_{\mu \nu}({\rm i} c\kappa)$ in Eqs.~(\ref{derexpa},\ref{derexpa2}) by its static limit $\alpha_{\mu \nu}(0) \equiv \alpha_{\mu \nu}^0$. Upon performing the $\xi$-integrals, we obtain the {\it retarded} Casimir-Polder potential
\begin{widetext}
\begin{align}
U_{\rm CP}& =-\frac{\hbar c}{ \pi d^4}\left\{\frac{\alpha_{\mu \mu}^0}{8} -\left(\frac{d}{R_1}+\frac{d}{R_2} \right) \left(\frac{3 \alpha_\perp^0}{40} +\frac{\alpha_{zz}^0}{15} \right)-\frac{1}{40}  \left(\frac{d}{R_1}-\frac{d}{R_2}\right) (\alpha_{xx}^0-\alpha_{yy}^0)+\frac{d^2}{30}\alpha_{zi}^0 \partial_i \left(\frac{1}{R_1}
+\frac{1}{R_2} \right)\right.\\\
&\left.+\left(\frac{d}{R_1}+\frac{d}{R_2} \right)^2 \!\left(\frac{3 \alpha_\perp^0}{280} -\frac{\alpha_{zz}^0}{240}\right) 
+\left[\left(\frac{d}{R_1}\right)^2\!+\left(\frac{d}{R_2} \right)^2 \right] \!\left(\frac{13 \alpha_\perp^0}{280}+\frac{3 \alpha_{zz}^0}{40}\right)\! +\frac{9}{560}\left[\left(\frac{d}{R_1}\right)^2\!-\left(\frac{d}{R_2} \right)^2  \right] (\alpha_{xx}^0-\alpha_{yy}^0) \right\}. \nonumber
\end{align}
\end{widetext}
In the special case of a spherical atom near a cylindrical metallic shell, the leading curvature correction in the above formula reproduces Eq.~(30) of Ref.~\cite{galina}.
Before  turning to the non-retarded limit, it is instructive to consider the classical high temperature limit,
where the  Casimir free energy is given by the first term of the Matsubara sum in Eq.~\eqref{eq.4}.
From the limit $\kappa\to 0$ of the coefficients $\beta^{(p)}_q$ we obtain the classical free energy as
\begin{widetext}
\begin{align}
U_{\rm classical}& = -\frac{k_B T}{2} \frac{1}{d^3} \left\{   \frac{1}{8} \alpha^0_\perp + \frac{1}{4} \alpha^0_{zz} 
-\frac{3}{64} \left( 3\frac{d}{R_1} + \frac{d}{R_2}\right) \alpha^0_{xx} 
-\frac{3}{64} \left( \frac{d}{R_1} + 3\frac{d}{R_2}\right) \alpha^0_{yy} 
-\frac{1}{16} \left(  \frac{d}{R_1}+\frac{d}{R_2}\right) \alpha^0_{zz}   \right. \\
&\!\!\!\!\!\!\! \left. +\frac{1}{128} \left( 17 \frac{d^2}{R_1^2} + 5 \frac{d^2}{R_2^2} +2 \frac{d^2}{R_1R_2}\right) \alpha^0_{xx}   
+\frac{1}{128} \left( 17 \frac{d^2}{R_2^2} + 5 \frac{d^2}{R_1^2} +2 \frac{d^2}{R_1R_2}\right) \alpha^0_{yy}
+\frac{1}{64} \left( 5\frac{d^2}{R_1^2} +5\frac{d^2}{R_2^2} -2\frac{d^2}{R_1R_2}\right) \alpha^0_{zz} \right\} \, . \nonumber
\end{align}
\end{widetext}
From the above result we obtain the {\it non-retarded} London interaction between the surface and a two-state atom for small distances $d\ll d_r = c/\omega_r$ at any finite temperature $T$. The dynamic dipolar polarizability of an atom or molecule on the imaginary frequency axis is given by 
\begin{equation}
  \label{eq:alpha_atom}
  \alpha_{\mu\nu}(\kappa) =  \frac{\alpha^0_{\mu\nu}}{1+(d_r \kappa)^2} \, .
\end{equation}
Formally, the non-retarded limit is obtained by taking the velocity of light to infinity ($c\to\infty$). This implies that the coefficients $\beta^{(p)}_q$ are evaluated at $\xi = \kappa_n d \sim 1/c \to 0$ while the atom's polarizability tends to the finite limit $\alpha_0/[1+(2\pi n k_B T/(\hbar \omega_r))^2]$ for $c\to\infty$. 
Hence the Matsubara sum over $n$ can be performed easily, leading to the non-retarded London potential at finite temperature $T$ of
\begin{equation}
  \label{eq:U_London}
  U_{\rm L} =  \frac{\hbar \omega_r}{2 k_B T} \coth\left( \frac{\hbar \omega_r}{2k_B T}\right)U_{\rm classical} \, .
\end{equation}

\section{Conclusions \& Outlook}

We have developed a derivative expansion for the Casimir-Polder potential between a small polarizable particle and a gently curved dielectric surface, which is valid in the limit of small particle-surface distances.  We have demonstrated the power of our approach by computing analytically the leading and next-to leading curvature corrections to the PFA for the potential, in the idealized limit of a perfectly conducting surface at zero temperature. For a two-level atom, we provide explicit formulae for the potential in the retarded Casimir-Polder limit, and in the non-retarded London limit.  

While the explicit results presented in the paper are for idealized situations, the gradient expansion method allows for many interesting
extensions: 
Specific dielectric properties of the surface can be easily incorporated and estimated numerically; resonances and anisotropy of the material
can lead to interesting interplay with shape and curvature.
On the side of the `atom' we can include effects from higher multipoles in the particle's polarizability. It is easy to deduce, already from
Eq.~(\ref{derexpa2}) that curvature of the surface can exert a torque, rotating an anisotropic particle into specific low energy orientation.
Non-equilibrium situations, involving an excited atom, or a surface held at a different temperature also provide additional avenues
for exploration.

\begin{acknowledgments}
We thank R.~L.~Jaffe  for valuable discussions.  This
research was supported by the NSF through grant No. DMR-12-06323.
\end{acknowledgments}

\appendix

\section*{Re-summation of the perturbative series}

 It has been recently shown that the derivative expansion of the Casimir energy between a flat and a curved surface follows from a resummation of  the perturbative series, for small in-plane momenta~\cite{fosco4}.  In this Appendix we show that the derivative expansion of the Casimir-Polder potential $U$ in Eq.~(\ref{derexpa}) can be justified by an analogous procedure. 
It is first convenient to recast the potential $U$ in Eq.~(\ref{eq.4}) in the form 
\be
U=-\frac{\hbar c}{d^4} \int_0^{\infty} \frac{d \xi}{2 \pi} \alpha_{\mu \nu}({\rm i} c\kappa)\;
U_{\mu \nu}(\xi)\;,\label{poten}
\ee
where the coefficients $U_{\mu \nu}$ depend linearly on the matrix elements of  ${\mathbb T}^{(S)}$. 
To specify the perturbative series, we introduce an arbitrary reference plane $\Sigma_2$
at distance $a$ from $\Sigma_1$ (see Fig.1), and  then  we set 
$H({\bf x})=a+h({\bf x})$. For sufficiently small $h$, the coefficients  $U_{\mu \nu}$ in Eq.~(\ref{poten}) admit the expansion: 
$$
U_{\mu \nu}=G_{\mu \nu}^{(0)}(a)+\sum_{n \ge 1} \frac{1}{n!}\int   d^2 {\bf x}_1 \dots  \int   d^2 {\bf x}_n
$$
\be \times\; G_{\mu \nu}^{(n)}(  {\bf x}_1,\cdots,  {\bf x}_n; a)\; h(  {\bf x}_1) \cdots h(  {\bf x}_n)\;,
\ee
where  $G_{\mu \nu}^{(0)}(a)$ denotes the coefficient for a planar surface at distance $a$ from the atom, $G^{(n)}_{\mu \nu}(  {\bf x}_1,\cdots,  {\bf x}_n; a)$ are  symmetric functions of $({\bf x}_1,\cdots ,{\bf x}_n)$,
and for brevity we have omitted the dependence of $G^{(n)}_{\mu \nu}$  on $\xi$. 
The  kernels $G^{(n)}_{\mu \nu}$  satisfy a set of differential relations, which result from    invariance of $U_{\mu \nu}$ under  a redefinition
of $a$ and $h({\bf x})$:
\be
a \rightarrow a+\epsilon\;,\;\;\;\;h({\bf x})\rightarrow h({\bf x})-\epsilon\;,\label{redef}
\ee
where $\epsilon$ is an arbitrary number. 
Independence of $U_{\mu \nu}$ on $\epsilon$ is equivalent to demanding  
${\partial^p U_{\mu \nu}}/{\partial \epsilon^p}|_{\epsilon=0}=0$
for all non-negative integers $p$. It is possible to verify that these conditions are satisfied if and only if
the kernels $G^{(n)}_{\mu \nu}$ obey the relations:
\begin{align}
& \frac{\partial^p G^{(n)}_{\mu \nu} }{\partial a^p}(  {\bf x}_1,\dots,  {\bf x}_n;a) \nonumber \\
&=\int d^2 {\bf x}_{n+1} \cdots \int d^2 {\bf x}_{n+p} G^{(n+p)}_{\mu \nu} (  {\bf x}_1,\dots,  {\bf x}_{n+p};a)\;.
\end{align}
In momentum space, the above relations read:
\be
\frac{\partial^p {\tilde G}^{(n)}_{\mu \nu} }{\partial a^p}(  {\bf k}_1,\dots,  {\bf k}_n; a)
=  {\tilde G}^{(n+p)}_{\mu \nu} (  {\bf k}_1,\dots,  {\bf k}_{n}, {\bf 0}, \dots ,{\bf 0};a)\;,\label{relat}
\ee
where our Fourier transforms are defined such that ${\tilde f}({\bf k})=\int d^2 {\bf x} f({\bf x}) \exp (-{\rm i} {\bf k} \cdot {\bf x})$, and we set ${\tilde G}^{(0)}_{\mu \nu} \equiv {G}^{(0)}_{\mu \nu} $. 
Consider now the perturbative expansion of the coefficients $U_{\mu \nu}$ in Fourier space:
\begin{align}
U_{\mu \nu}&=G_{\mu \nu}^{(0)}(a)+\sum_{n \ge 1}\frac{1}{n!}\int \frac{d^2{\bf k}_1}{4 \pi^2} \cdots \int   \frac{d^2{\bf k}_n}{4 \pi^2}  \nonumber\\
& \times\; {\tilde G}_{\mu \nu}^{(n)*}(  {\bf k}_1,\cdots,  {\bf k}_n; a)\; {\tilde h}(  {\bf k}_1) \cdots {\tilde h}(  {\bf k}_n)\;,
\end{align}
For profiles of small slopes  ${\tilde h}({\bf k})$ is supported near zero, and then it is  legitimate to Taylor-expand ${\tilde G}^{(n)}({\bf k}_1,\cdots ,{\bf k}_n)$   in powers of the in-plane momenta $({\bf k}_1, \cdots ,{\bf k}_n)$. Upon truncating the Taylor expansion to fourth order, and after going back to position space, we find for $U_{\mu \nu}$ the expression 
\begin{widetext}
\begin{align}
 U_{\mu \nu} & \simeq G_{\mu \nu}^{(0)}(a)+\sum_{n \ge 1}\left\{ \frac{1}{n!} A^{(n)*}_{\mu \nu}(a) h^n({\bf 0} )+\frac{h^{n-1}({\bf 0} )}{(n-1)!}\left[-\frac{1}{2}B^{(n)*}_{\mu \nu | i j}(a) 
 \partial_i \partial_j h({\bf 0})
+\frac{{\rm i}}{3!}B^{(n)*}_{\mu \nu | i j k}(a)  \partial_i \partial_j \partial_k h({\bf 0}) \right. \right. \nonumber\\
& \left.\left.+
\frac{1}{4!}B^{(n)*}_{\mu \nu | i j k l} (a)\partial_i \partial_j \partial_k \partial_l h({\bf 0})
\right]\right\}+\sum_{n \ge 2}\frac{h^{n-2}({\bf 0} )}{8 (n-2)!}C^{(n)*}_{\mu \nu | i j k l} (a) \partial_i \partial_j h({\bf 0})\partial_k \partial_l h({\bf 0})\;,\label{infsum}  \, 
\end{align}
\end{widetext} 
where 
\be A^{(n)}_{\mu \nu}(a)={\tilde G}_{\mu \nu}^{(n)}( {\bf 0},\cdots,{\bf 0};a)\;,\ee 
\be
B^{(n)}_{\mu \nu | i_1 \dots i_p}(a)= \partial_{k_{i_1}}c\cdots \partial_{k_{i_p}} {\tilde G}_{\mu \nu}^{(n)}({\bf k}, {\bf 0},\cdots,{\bf 0};a) \vert_{{\bf k}=0}\;,\label{defB}
\ee
and
\be
C^{(n)}_{\mu \nu | i j k l}(a)= \partial_{k_{i}} \partial_{k_{j}} \partial_{k'_k} \partial_{k'_l} {\tilde G}_{\mu \nu}^{(n)}({\bf k},{\bf k}', {\bf 0},\cdots,{\bf 0};a) \vert_{{\bf k}={\bf k}'=0}\;,\label{defC}
\ee
and we have only displayed the terms that do not vanish identically on account of the condition  $\nabla h({\bf 0})=0$. 
The $n$-sums in Eq.~(\ref{infsum}) can be easily done, because by virtue of Eq.~(\ref{relat}) the $A, B, C$ coefficients satisfy the relations:
\be
A^{(n)}_{\mu \nu}(a)=\frac{\partial^n G^{(0)}_{\mu \nu}}{\partial a^n}\;,
\ee
\be
B^{(n)}_{\mu \nu | i_1 \dots i_p}(a)=\frac{\partial^{n-1}  B^{(1)}_{\mu \nu | i_1 \dots i_p}(a)}{\partial a^{n-1}}\;,\label{Bc}
\ee
and
\be
C^{(n)}_{\mu \nu | i j k l}(a)=\frac{\partial^{n-2}  C^{(2)}_{\mu \nu | i j k l }(a)}{\partial a^{n-2}}\;.\label{Cc}
\ee
After we substitute the above relations into Eq.~(\ref{infsum}), and recalling that $d=a+h({\bf 0})$, we obtain the desired result:
\begin{widetext}
\be
U_{\mu \nu} \simeq G_{\mu \nu}^{(0)}(d)-\frac{1}{2}B^{(1)*}_{\mu \nu | i j}(d) 
 \partial_i \partial_j h({\bf 0})
+\frac{{\rm i}}{3!}B^{(1)*}_{\mu \nu | i j k}(d)  \partial_i \partial_j \partial_k h({\bf 0}) +
\frac{1}{4!}B^{(1)*}_{\mu \nu | i j k l} (d)\partial_i \partial_j \partial_k \partial_l h({\bf 0})+\frac{1}{8}C^{(2)*}_{\mu \nu | i j k l} (d) \partial_i \partial_j h({\bf 0})\partial_k \partial_l h({\bf 0})\;.
\ee
\end{widetext}
We see that the re-summed perturbative series involves the coefficients $B^{(1)}_{\mu \nu | i_1\dots i_p}(d)$, $p=2,3,4$ and $C^{(2)}_{\mu \nu | i j k l} (d)$, evaluated for $a=d$. As  is apparent from Eqs.~(\ref{defB}-\ref{defC}), these coefficients can be extracted, respectively, from the first and second order kernels ${\tilde G}^{(1)}_{\mu \nu}({\bf k};d)$ and ${\tilde G}^{(2)}_{\mu \nu}({\bf k}_1,{\bf k}_2;d)$, by Taylor-expanding them for small momenta.

\end{document}